\documentclass{PoS}

\title{GMRT study of $X$-shaped radio sources}

\ShortTitle{GMRT study of $X$-shaped radio sources}

\author{\speaker{Dharam Vir Lal}%
         \thanks{We thank the staff of the GMRT who have made
these observations possible.
GMRT is run by the National Centre for Radio Astrophysics
of the Tata Institute of Fundamental Research..}\\
        Max-Planck-Institut f\"ur Radioastronomie,
        Auf dem H\"ugel 69, 53121 Bonn, Germany\\
        E-mail: \email{dharam@mpifr-bonn.mpg.de}}

\author{A. P. Rao\\
        National Centre for Radio Astrophysics (TIFR),
        Pune University Campus, Pune 411007, India}

\author{Martin J. Hardcastle\\
        School of Physics, Astronomy, and Mathematics, University of Hertfordshire, Hatfield, UK}

\author{C. C. Cheung \& Sanjay Bhatnagar\\
        National Radio Astronomy Observatory, Socorro, P.O. Box 0, New Mexico, USA}
\author{Ralph P. Kraft\\
        Harvard-Smithsonian Center for Astrophysics, MS-67, Cambridge, MA 02138, USA}

\author{Andrei P. Lobanov \& Anton J. Zensus\\
        Max-Planck-Institut f\"ur Radioastronomie,
        Auf dem H\"ugel 69, 53121 Bonn, Germany}

\abstract{{\em Context.} The nature of $X$-shaped sources
is a matter of considerable
debate in the literature: it has even been proposed that they provide
evidence for black-hole-mergers$/$spin-reorientation, and therefore
constrain the rate of strong gravitational wave events. \\
{\em Aim.} To explore the nature of these $X$-shaped
radio galaxies. \\
{\em Method.} We conduct a systematic study of a large sample
of known and newly discovered $X$-shaped sources
along with a comparison sample.  We used the Giant Metrewave
Radio Telescope with
resolution of $\sim$6$^{\prime\prime}$ to $\sim$15$^{\prime\prime}$
at 610 MHz and 240 MHz in the dual-frequency mode. \\
{\em Preliminary Result.}  Based on our careful analysis and
estimation of the possible systematic errors, the known $X$-shaped sources
divide into the following three categories:
(i) the wings have flatter spectral indices than the active lobes,
(ii) the wings and the active lobes have comparable spectral indices, and
(iii) the wings have steeper spectral indices than the active lobes.
In addition, based on our preliminary
analysis, one sample source from our comparison sample shows
a spectral index result belonging to category (i). \\
{\em Future.} Milliarcsecond scale imaging will be conducted on some
of these $X$-shaped sources to investigate if they are examples of
binary AGN systems and thereby understand the nature of these sources.}

\FullConference{From Planets to Dark Energy: the Modern Radio Universe\\
                 October 1-5 2007\\
                 The University of Manchester, UK}

\begin{document}

\section{Introduction}

\noindent
A peculiar and small subclass of extragalactic radio sources
called $X$-shaped, or `winged' sources
are characterised by two low-surface-brightness
lobes (the `wings') oriented at an angle to the `active', or
high surface brightness radio lobes, giving the total source
an `$X$' shape. These two sets of lobes usually pass symmetrically through
the centre of the associated host galaxy.
Merritt \& Ekers (2002) noted that the majority of these sources
are of Fanaroff-Riley type II (FR~II) (Fanaroff \& Riley 1974)
and the rest are either FR~I or mixed.

\noindent {\bf Formation Scenario}
Several authors have attempted to explain the unusual
structure in $X$-shaped sources.
These $X$-shaped radio sources have been put forth as
derivatives of central engines that have been reoriented,
perhaps due to a minor merger (Merritt \& Ekers 2002;
Dennett-Thorpe et al. 2002; Gopal-Krishna et al. 2003).
Alternatively, they may also result from two pairs of jets,
which are associated with a pair of unresolved AGNs (Lal \& Rao 2005, 2007).
These, however, are not the only interpretations for the unusual morphologies;
some authors suggest a hydrodynamic origin
(Leahy \& Williams 1984; Worrall et al. 1995; Capetti et al. 2002;
Kraft et al. 2005) and some suggest a conical precession
of the jet axis (Rees 1978; Parma et al. 1985; Mack et al. 1994).
See Lal \& Rao (2007) and Cheung (2007) for a detailed account.

\section{Sample \& GMRT Observations}

\noindent {\bf Known sample}
The earlier sample of known $X$-shaped sources was drawn from the list
mentioned in Merritt \& Ekers (2002) compiled by Leahy \& Parma (1992).
%There are nearly a dozen such sources,
These source have been selected solely on the basis of their morphology,
and the sample is inhomogeneous and in no sense a statistical complete sample.

\noindent {\bf Comparison sample}
The comparison sample consists of all nearby ($z < 0.1$)
normal FR~II sources from the 3CRR catalogue. These sources have radio
luminosities similar to that of the $X$-shaped sources, which lie close
to the FR~I/FR~II divide.  We impose an angular-size cutoff (based on
high-frequency radio maps) on the target sample and ensure that our
sample sources are of similar angular size to typical $X$-shaped sources.
In addtion, the sample sources have known weak transverse extensions
(proto-wings?) and also have X-ray ({\it XMM}/{\it Chandra}) observations.

\noindent {\bf New sample}
The new sample is drawn from the compiled list
of nearly 100 new candidate $X$-shaped radio sources through a search of the
FIRST survey database (Cheung 2007).  Our sample sources had
(i)~characteristic `$X$' shape,
(ii) both set of lobes passing symmetrically through the
centre of the associated host galaxy, and
(iii) an angular size of more than 1.2$^{\prime}$ as seen in the VLA--FIRST
1.4~GHz images. \\ [-0.2cm]

\noindent {\bf GMRT Observations}
The 240 MHz and 610~MHz feeds of GMRT (Swarup et~al. 1991) are
coaxial feeds and therefore, simultaneous multi-frequency observations
at these two frequencies are possible.
We made synthesis observations of all our sample sources at
240 MHz and 610 MHz, in the dual frequency mode, using the GMRT
during several observing GTAC cycles, in the standard spectral line mode.
%
%The full synthesis radio images shown in Figs.~\ref{full_syn1}
%and \ref{full_syn2}
%have nearly complete UV coverage, an angular resolution
%$\sim$14$^{\prime\prime}$ and $\sim$6$^{\prime\prime}$ and the
%rms~noise in the maps are $\sim$2.0 and
%$\sim$0.2 mJy~beam$^{-1}$ at 240 and 610~MHz, respectively.
%The dynamic ranges in the two maps are
%$\sim$1600 and $\sim$4700 in case of 3C~52 and
%$\sim$200 and $\sim$600 in case of B1059$+$169,
%respectively at 240 and 610~MHz.
The GMRT has a hybrid configuration (Swarup et~al. 1991)
with 14 of its 30 antennas located in a central compact array
with size $\sim$1.1~km and the remaining antennas distributed in
a roughly `Y' shaped configuration, giving a maximum baseline length
of $\sim$25~km.
The hybrid configuration gives reasonably good sensitivity
for both compact and extended sources.

\section{Preliminary Results}

\noindent
The spectral characteristics of known $X$-shaped sources
seem to fall into three distinct categories, namely, sources in which
(i) the wings have flatter spectral indices than the active lobes,
(ii) the wings and the active lobes have comparable spectral indices, and
(ii) the wings have steeper spectral indices than the active lobes.
While it is probable that the three categories
of sources are unrelated to one another,
a single model to explain these sources is a challenge (Lal \& Rao 2007).

We find evidence for the presence of possible unusual spectral
properties in typical FR~II radio galaxies; i.e., from an independent
study of our comparison sample of FR~II radio sources,
one source from our sample seems to have relatively flatter spectral
indices for the low-surface-brightness features than the
high-surface-brightness features.

    \parbox[c]{0.30\textwidth}{\includegraphics[width=0.30\textwidth]{B1059_610.PS}}
    \hfill
    \parbox[c]{0.30\textwidth}{\includegraphics[width=0.30\textwidth]{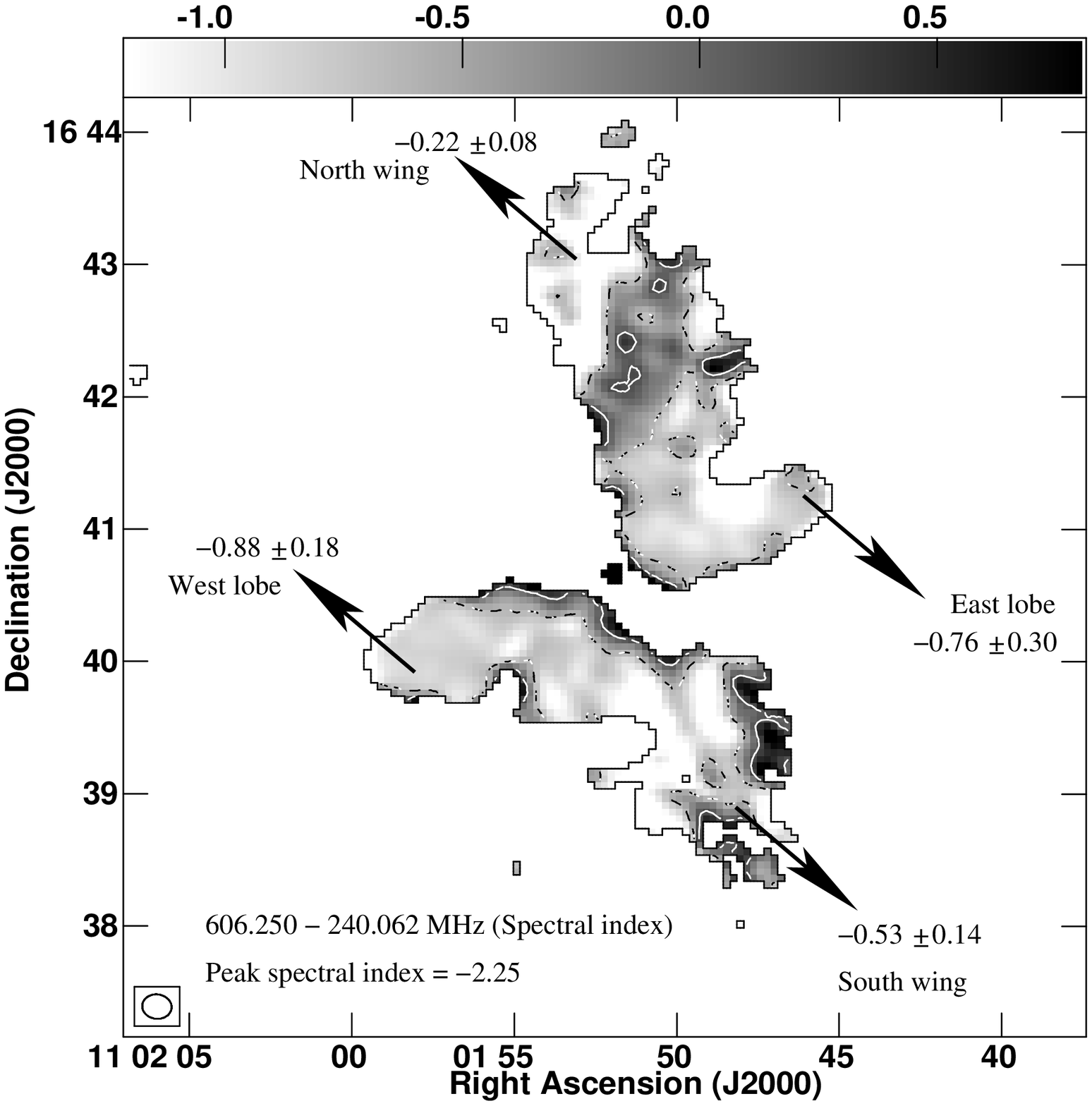}}
    \hfill
    \parbox[c]{0.30\textwidth}{\includegraphics[width=0.30\textwidth]{B1059_240.PS}} \\

    \parbox[c]{0.30\textwidth}{\includegraphics[angle=-90,width=0.30\textwidth]{3C452_RR.PS}}
    \hfill
    \parbox[c]{0.30\textwidth}{\includegraphics[angle=-90,width=0.30\textwidth]{3C452_SPIX.PS}}
    \hfill
    \parbox[c]{0.30\textwidth}{\includegraphics[angle=-90,width=0.30\textwidth]{3C452_LL.PS}} \\ [0.2cm]
{\bf Figure} GMRT map of B1059$+$169 (a known $X$-shaped source)
at 610 MHz (top left panel), 240~MHz (top right panel) and the spectral index
(240~MHz, 610~MHz) map (top middle panel).
Similarly, the lower panel shows the GMRT map of 4C 452 (a FR~II source,
possibly missed from the $X$-shaped sample due to projection),
from our comparison sample
at 610 MHz (bottom left panel), 240~MHz (bottom right panel) and
the spectral index (240~MHz, 610~MHz) map (bottom middle panel).
The CLEAN beams for 610 MHz and 240 MHz maps are
$\sim$6$^{\prime\prime}$ and $\sim$15$^{\prime\prime}$, respectively.
The GMRT map at 610~MHz is matched with the resolution of 240~MHz.
Note the unusual spectral index in B1059$+$169.

%\section*{Important References}
%
%Kraft, R.P., Hardcastle, M.J., Worrall, D.M. \& Murray, S.S. 2005, ApJ, 622, 149 \\
%Lal, D.V. \& Rao, A.P. 2007, MNRAS, 374, 1085 \\
%Merritt, D. \& Ekers, R.D. 2002, Sci, 297, 1310
%

\end{document}